\documentclass{iopart}

\usepackage{color}
\usepackage{graphicx,epsfig}
\usepackage{bm}
\usepackage{amssymb}

\gdef\ffrac#1#2{\textstyle{#1\over#2}\displaystyle}
\def\zb{{\bar z}}

\begin{document}
\title[The Ubiquitous `$c$']
{The Ubiquitous `$c$': from the Stefan-Boltzmann Law to Quantum
Information\footnote{Boltzmann Medal Lecture, Statphys24, Cairns,
July 2010}}
\author{John Cardy$^1$}
\address{$^1$Oxford University, Rudolf Peierls Centre for
      Theoretical Physics, 1 Keble Road, Oxford, OX1 3NP, United Kingdom
      and All Souls College, Oxford}

\date{\today}

\begin{abstract}
I discuss various aspects of the role of the conformal anomaly
number $c$ in 2- and 1+1-dimensional critical behaviour: its
appearance as the analogue of Stefan's constant, its fundamental
role in conformal field theory, in the classification of 2d
universality classes, and as a measure of quantum entanglement,
among other topics.
\end{abstract}

\maketitle

\section{The Stefan-Boltzmann Law}
It seems appropriate to honour the founder of our subject, after
whom this award is named, by recalling one of his most elegant
pieces of work. In 1879 Josef Stefan had published his famous law
\cite{Stefan} stating that the power per unit area radiated by an
ideal black body is proportional to the fourth power of its
absolute temperature $T$. In terms of the energy density $u$ in
the radiation field, this may be written as
$$
u=(4\pi/c)\,\sigma T^4\,,
$$
where $\sigma$ is Stefan's constant, and $c$ here is not the
conformal anomaly number of my title, but the (even more
ubiquitous) speed of light. Stefan apparently based his empirical
law on the analysis of experimental data of John Tyndall
\cite{Tyndall}, a Victorian scientist now more famous for his work
in magnetism as well as glaciology\footnote{As well as Alpine
mountaineering -- among many famous climbs he made the first
ascent of the Weisshorn in the Swiss Alps. The strong correlation
between physicists and climbers was evident even at that time.}.

Stefan was Boltzmann's advisor \cite{Bbiog}, and it must have been
a great pleasure to him when his former student produced
\cite{Boltzmann} a theoretical derivation of this law based on
thermodynamic reasoning. It is worth recalling Boltzmann's
derivation, as it illustrates the power, as well as the potential
pitfalls, of combining ideas from two different branches of
physics. He imagined a box, filled with black-body radiation, of
which one wall is a piston which is moved by the pressure of the
radiation. According to classical electromagnetism, the pressure
$P$ is related to the energy density by $P=\frac13u$. The
heat-energy balance equation then reads
$$
TdS=d(uV)+PdV=u'(T)VdT+udV+\ffrac13udV=u'(T)VdT+\ffrac43udV\,.
$$
Dividing through by $T$ and using the fact that $dS$ is a complete
differential,
$$
\frac\partial{\partial V}\left(\frac{u'(T)V}T\right) =
\frac\partial{\partial T}\left(\frac{4u}{3T}\right)\,,
$$
which simplifies, after some algebra, to $u'(T)=4u(T)/T$, that is,
$u(T)\propto T^4$.\footnote{Note that in $d$ space dimensions an
analogous argument leads to $u(T)\propto T^{d+1}$.}

This simple yet elegant argument was praised by Lorentz in 1907
\cite{Lorentz} as `a jewel of theoretical physics.' But, with
hindsight, we see that Boltzmann was either fortunate, or, more
likely, inspired, in juxtaposing results of classical
electromagnetism with the idea that radiation behaves like a
fluid. For the immediate question is then ``a fluid of what kind
of particle?'' -- and of course this was not answered, even
heuristically, until Planck made his hypothesis, and, more
systematically, until the quantization of the radiation field.
Indeed, as every student knows, $\sigma=\pi^4k^4/60\hbar^3c^2$,
and, once again with hindsight, a little dimensional analysis
would have shown Boltzmann that, if Stefan's constant depends on
fundamental constants, one of these must contain the dimensions of
mass and therefore be something which at the time was unknown in
the classical physics of the pure radiation field.

In more modern terms, the equation $P=\frac13u$, which was the
starting point of Boltzmann's argument, is equivalent to the
statement that the energy-momentum tensor $T_{\mu\nu}$ is
traceless:
$$
u-3P=T_{00}-\sum_{j=1}^3T_{jj}=T^\mu_\mu=0\,.
$$
This equation is true for the classical Maxwell field, but
Boltzmann was implicitly assuming that it also holds for the
quantized field. Yet nowadays we know many examples of field
theories for which the energy-momentum tensor is traceless at the
classical level, but not when the theory is quantized properly.
Examples are electrodynamics coupled to (massless) particles, with
non-trivial vacuum polarization effects (so in fact Stefan's law
does not hold in QED at high temperatures), and non-Abelian gauge
theories.

In fact it is now understood that in a quantum field theory
$$
T^\mu_\mu\propto \beta(g)\,,
$$
where $\beta(g)$ is the renormalization group (RG)
beta-function\footnote{The appearance of a non-zero value for the
trace after quantization is often (confusingly) referred to as the
`conformal anomaly'. However this is not the anomaly related to
$c$ which is the object of the subsequent discussion.}. This means
that, even when fluctuations (either quantum or thermal) are taken
into account, the trace $T_\mu^\mu$ vanishes at RG fixed points,
that is, in a critical theory at which the correlation length
diverges.

Thus we expect a generalized Stefan-Boltzmann law to hold at all
quantum critical points which have a relativistic dispersion law
$E\sim v|k|$ at low energies, where, however, $v$ does not have to
be the speed of light, but could, for example, be the Fermi
velocity or the speed of sound. In 1+1 dimensions, with which we
shall henceforth be concerned, there are many such examples: free
fermions at finite density, Luttinger liquids, quantum Hall edge
states, and many critical quantum spin chains. In that case, the
1+1-dimensional analogue of the Stefan-Boltzmann law takes the
form
\begin{equation}\label{SB}
 u=\frac{\pi c}{6\hbar v}\,(kT)^2\,,
\end{equation}
where now $c$ is not the speed of light, but, as a simple
calculation in quantum statistical mechanics shows
$$
c=\left\{ \begin{array}{r@{\quad:\quad}l}1 & \mbox{single free
boson;} \\ N & \mbox{$N$ free bosons;} \\
\ffrac12 & \mbox{a spinless fermion.}\end{array}\right.
$$
However, in general, $c$ is fractional, and indeed is the first
appearance of the ubiquitous conformal anomaly number.

\section{Why is the conformal anomaly anomalous?}
In order to understand how $c$ arises in conformal field theory,
let us consider the simplest possible example of a single scalar
field $\phi(r)$ with action
$$
S=\int(\nabla\phi)^2d^2\!x\,.
$$
In 2 space dimensions, this might represent the energy in the
electrostatic field, or of spin waves at low temperatures. Many
important models of statistical physics in 2 or 1+1 dimensions can
be `bosonized' into this simple form. In order to understand why
this is conformally invariant, it is useful to define so-called
complex coordinates
$$
z=x_1+ix_2\qquad \zb=x_1-ix_2\,,
$$
in terms of which
$$
S\propto\int(\partial_z\phi)(\partial_\zb\phi)d^2\!z\,.
$$
In two dimensions conformal mappings correspond locally to
analytic functions $z\to z'=f(z)$, and we can see that,
classically, $S$ is indeed invariant under these, since
$\partial_z=f'(z)\partial_{z'}$,
$\partial_\zb=\overline{f'(z)}\partial_{\zb'}$ and
$d^2\!z=|f'(z)|^{-2}d^2\!z'$. Indeed, the trace $T^\mu_\mu$, as
calculated by Noether's theorem, vanishes identically. The
non-zero components, in complex coordinates, are
$$
T=T_{zz}=-(\partial_z\phi)^2\quad\mbox{and}\quad \overline
T=T_{\zb\zb}=-(\partial_\zb\phi)^2\,.
$$
Under a conformal mapping $z\to z'=f(z)$, we see that classically
$T$ transforms simply:
$$
T(z)=\big(f'(z)\big)^2\,T(z')\,,
$$
and similarly for $\overline T$.

However, once the fluctuations are taken into account, this is no
longer the case. In fact $T$ as defined above is divergent, since
$\langle\phi_k\phi_{-k}\rangle\propto 1/k^2$, so $\langle
T\rangle\sim \int k^2(d^2k/k^2)$. One way to define it properly is
by point-splitting:
$$
T(z)=\lim_{\delta\to0}\left[\partial_z\phi(z+\ffrac\delta
2)\partial_z\phi(z-\ffrac\delta 2)-\frac1{2\delta^2}\right]
$$
where $\delta$ is a short-distance cut-off of the order of the
lattice spacing. However this subtraction does not in general
commute with conformal mappings, because in the $z'$-plane we
should still subtract off $1/2\delta^2$ rather than
$1/2(|f'(z)|\delta)^2$. The result is the appearance of an
anomalous term in the transformation law for $T$:
\begin{equation}\label{T}
T(z)=f'(z)^2\,T(f(z))-\frac c{12}\{f,z\}\,,
\end{equation}
where $\{f,z\}=(f'''f'-\ffrac32{f''}^2)/{f'}^2$ (the Schwartzian
derivative), and in this case $c=1$. This is a classic example of
the appearance of an anomaly in quantum field theory, when a
symmetry (in this case, conformal symmetry) is not respected by
the necessary regularization procedure. In fact the form of the
last term on the rhs of (\ref{T}), although complicated, is
completely fixed by the requirement that it hold under a general
iterated sequence of conformal mappings. The only arbitrariness is
in the coefficient $c$, which is thereby a fixed parameter
characterizing the particular CFT or universality class.

Although (\ref{T}) might appear rather technical, from this result
in fact flow all the various ubiquitous physical manifestations of
$c$.

\section{Stefan-Boltzmann in 1+1 dimensions}
\begin{figure}[h]
\centering
\includegraphics[width=0.5\textwidth]{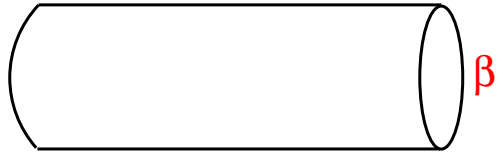}
\caption{The finite temperature partition function is given by the
path integral on a cylinder.}\label{fig1}
\end{figure}
Consider a critical 1+1-dimensional quantum system (with a linear
dispersion relation $E\sim v|k|$ at low energies) of length $L$,
at low but finite temperature $T$. As Feynman taught us, the
partition function ${\rm Tr}\,e^{-\beta\hat H}$ is given by the
path integral in imaginary time with periodic boundary conditions
modulo $\beta\hbar$, where $\beta=1/kT$. That is, it is equivalent
to a 2-dimensional classical system on a cylinder (see
Fig.~\ref{fig1}) of circumference $\beta\hbar$. This is
conformally related to the plane by the mapping $z\to
z'=(\beta\hbar/2\pi)\log z$. Applying (\ref{T}) and using the fact
that $\langle T\rangle_{\rm plane}=0$, we find \cite{BCN,Affleck}
the result for the energy density
\begin{equation}\label{u}
u=\langle T_{00}\rangle=(1/2\pi)\big(\langle T\rangle_{\rm
cylinder}+\langle\overline T\rangle_{\rm cylinder}\big)=\frac{\pi
c}{6\hbar}\,(kT)^2\,,
\end{equation}
the 1+1-dimensional version (\ref{SB}) of the Stefan-Boltzmann
law, in units where $v=1$. Note that this corresponds to a linear
specific heat. In principle, this can be compared with experiment,
although this requires a separate determination of $v$.

Equivalently, we can think of the coordinate along the cylinder as
representing imaginary time, in which case we have a
1+1-dimensional quantum theory defined on a finite spatial
interval $\ell=\beta\hbar$, with periodic boundary conditions. In
that case (\ref{u}) gives the finite-size corrections to the
ground state energy, $E_0\sim O(\ell)-(\pi c/6\ell)$, once again
in units where $v=1$. This is one the most effective ways of
determining $c$ from numerical or analytic diagonalization of the
hamiltonian $\hat H$.

\section{The conformal periodic table}
As with any quantum theory, it is advantageous to realise the
symmetries of CFT in terms of generators acting on the Hilbert
space of states of the theory. In this case these are the
so-called Virasoro generators $L_n=(1/2\pi i)\oint z^{n+1}T(z)dz$
(and analogously $\overline L_n$). The transformation law
(\ref{T}) is completely equivalent to the Virasoro algebra
$$
[L_n,L_m]=(n-m)L_{n+m}+\ffrac1{12}c\,n(n^2-1)\delta_{n,-m}\,.
$$
In particular $L_0$ and $\overline L_0$ generate scale
transformations and rotations, and the scaling fields of the CFT
correspond to eigenstates of these operators, with eigenvalues
giving all the critical exponents. Thus the study of the
representation theory of the Virasoro algebra gives a way of
classifying all possible CFTs and thereby universality classes in
2d. The breakthrough in this direction came following the seminal
1984 paper of Belavin, Polyakov and Zamoldchikov (BPZ) \cite{BPZ}
in which they showed that, for certain special rational values of
$c<1$, the CFT closes with only a finite number of representations
of the Virasoro algebra, and, for, these cases, all the critical
exponents and multi-point correlation functions are calculable.
Shortly thereafter Friedan, Qiu and Shenker \cite{FQS} showed that
unitary CFTs (corresponding to local, positive definite Boltzmann
weights) are a subset of this list, with $c=1-6/m(m+1)$ and $m$ an
integer $\geq3$. This gives rise to what might be termed the
`conformal periodic table' (Table~\ref{PT}).
\begin{table}[h]
\centering
\begin{tabular}{|c|l|l|}\hline
$c$ & Scaling dimensions & Universality class\\
\hline\hline $\frac12$ & $0,\frac18,1$ & Critical Ising\\ \hline
$\frac7{10}$ &
$0,\frac3{80},\frac1{10},\frac7{16},\frac35,\frac32$ & Tricritical
Ising\\ \hline $\frac45$ &
$0,\frac1{40},\frac1{15},\frac18,\frac25,\frac{21}{40},\frac23,\frac75,\frac{13}8,3$
& Tetracritical Ising\\ \hline $\frac45$ &
$0,\frac1{15}\times2,\frac25,\frac23\times2,\frac75,3$ & Critical
3-state Potts\\ \hline $\vdots$ & $\vdots$ & $\vdots$\\ \hline
\end{tabular}
\caption{The first few elements in the conformal periodic
table.}\label{PT}
\end{table}
The first few examples may be identified with well-known
universality classes. The `hydrogen atom' of CFT is the scaling
limit of the critical Ising model, `helium' is the tricritical
Ising model, and so on. Note, however, that at the next value of
$c=\frac45$ two possible `isotopes' arise. In the second,
corresponding to the critical 3-state Potts model, not all the
scaling dimensions allowed by BPZ in fact occur, but some of those
that do actually appear twice. In fact the constraint of unitarity
is not sufficient to determine exactly which representations
actually occur in a given CFT. The answer to this is provided by
demanding consistency of the theory on a torus \cite{JCmod}, by
interchanging the intepretations of space and imaginary time
similar to the case of the cylinder mentioned above. For the
torus, this is a modular transformation, and the requirement of
modular invariance has become another powerful tool in classifying
CFTs, completely solved in the case $c<1$ by Cappelli, Itzykson
and Zuber \cite{CIZ}.

\section{$c$ and entanglement entropy}
In recent years, $c$ has been playing a new role in quantifying
the degree of quantum entanglement in the ground state of a
1+1-dimensional critical system (for example, a quantum spin
chain.) There is by now a huge literature on this \cite{review},
but here I will concentrate on the simplest possible scenario: a
long system, of length $L$, near a quantum critical point which is
divided into two halves $A$ and $B$ at the origin. We assume that
the degrees of freedom in $A$ are accessible only to observer A,
and conversely those in $B$ only to B. In the ground state, A's
observations are entangled with those of B: if A performs a
measurement with a certain outcome, this can restrict the possible
outcomes of measurements B can subsequently make. A very useful
way to measure the entanglement between $A$ and $B$ is through the
A's reduced density matrix
$$
\rho_A={\rm Tr}_B\,|0\rangle\langle0|\,,
$$
and the so-called R\'enyi entropies
$$
S_A^{(n)}=(1-n)^{-1}\log{\rm Tr}_A\,\rho_A^n\,.
$$
The limit $n\to1$ gives the well-known von Neumann entropy $-{\rm
Tr}_A\rho_A\log\rho_A$. If the state $|0\rangle$ is unentangled,
$\rho_A$ has a single eigenvalue $=1$, so that $S_A^{(n)}=0$.
However, a maximally entangled state will give rise to $O(e^{{\rm
const.}L})$ eigenvalues all of the same order, so that
$S_A^{(n)}=O(L)$.

\begin{figure}[h]
\centering
\includegraphics[width=0.5\textwidth]{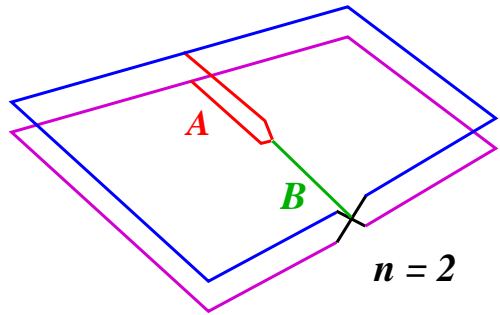}
\caption{The surface on which the path integral giving the R\'enyi
entropy for $n=2$ is evaluated.}\label{fig2}
\end{figure}

It turns out that ${\rm Tr}\,\rho_A^n$ is given by the path
integral, or partition function, on an $n$-sheeted surface ${\cal
R}_n$ with a branch cut running from the origin to the boundary,
as shown for the case $n=2$ in Fig.~\ref{fig2}.
This however, is related the full plane by the conformal mapping
$z\to z'=z^{1/n}$. Using once again (\ref{T}), we then find
$$
\langle T\rangle_{{\cal R}_n}=\frac{c(1-n^{-2})}{12z^2}\,.
$$
This behaviour means that the branch point behaves like the
insertion of a scaling operator with dimension
$x_n=(c/12)(n-n^{-1})$, and thus the partition function on ${\cal
R}_n$ goes like $L^{-x_n}$ at the critical point, or $\xi^{-x_n}$
away from it, where $\xi$ is the correlation length. Thus
$$
S_A^{(n)}\sim \frac{c}{12}(1+n^{-1})\,\log L\,,
$$
for $L\gg\xi$, with $L$ replaced by $\xi$ in the opposite limit.
This result \cite{Holzhey,cc1} has been verified by numerous
analytic computations in exactly solvable models, and has become
the gold standard for measuring $c$ numerically, using, for
example, density matrix RG methods. However, it is but the tip of
the iceberg in results of this type: taking, for example, $A$ to
consist of two disjoint intervals gives access to the entire
spectrum of scaling dimensions of the CFT, as well as the operator
product expansion coefficients \cite{review}.

\section{Other appearances of $c$}
These are too numerous to list exhaustively, but some of my
favourites are:
\begin{itemize}
\item if we put a CFT on a manifold of Euler character $\chi$ and linear size $L$,
the free energy has the asymptotic form as $L\to \infty$
\cite{CardyPeschel}
$$
F\sim AL^2 +BL-\ffrac16{c}\chi\log L+\cdots\,.
$$
This works even for a disc, with $\chi=1$. If there are corners on
the boundary, however, the coefficient is modified in a known
manner.
\item one cannot give a survey of the role of $c$ without
mentioning Zamolodchikov's beautiful $c$-theorem \cite{Zam}: \em
There exists a function $C(g)$ of the coupling constants $\{g\}$
which is decreasing along RG flows and is stationary at RG fixed
points, where it equals $c$\em. This implies that RG flows, at
least in 2d, go `downhill' and rules out (at least for unitary
theories) exotic behaviour such as limit cycles. Remarkably, no
analogous result has been shown in higher dimensions (except in
the case of a high degree of supersymmetry), and it remains an
open question as to whether it is in fact true.
\item one nice consequence of the above is the $c$-theorem sum rule \cite{JCc}: suppose
that we move slightly away from the critical point by adding, for
example, a magnetic field $H$. Then in that case the $q$-dependent
susceptibility $\chi(q)$ is analytic at wave vector $q=0$, and its
curvature there, by scaling, goes like $H^{-2}$. In fact the
coefficient is universal:
$$
{c}=3\pi^2\left(\frac\delta{\delta+1}\frac
H{k_BT_c}\right)^2\,\left.\chi''(q)\right|_{q=0}\,,
$$
where $\delta$ is the usual critical exponent in $M\sim
H^{1/\delta}$.
\end{itemize}
This is only a very selective list. Indeed, one may truthfully say
that, at least in two or 1+1 dimensions, $c$ is everywhere!

\section*{Acknowledgments}
I am immeasurably grateful to all the collaborators and other
colleagues who have supported my research over the years, many of
whom have played no small part in achieving many of the results
described above. I thank the members of the IUPAP C3 Commission on
Statistical Physics for the award of the Boltzmann Medal on this
occasion, which I consider to be a recognition of the whole field
theoretic approach to critical behaviour and the work of all the
people who have contributed to this effort.

\end{document}